\documentclass[12pt]{article}
\usepackage{epsfig}

\def\e{\begin{equation}}
\def\ee{\begin{eqnarray}}
\def\ff{\end{eqnarray}}
\def\f{\end{equation}}
\def\=#1{\overline{\overline #1}}
\def\_#1{{\bf #1}}

\def\E{\epsilon}
\def\M{\mu}

\def\d{\partial}
\def\.{\cdot}

\def\^{\hat}

\def\l#1{\label{eq:#1}}
\def\r#1{(\ref{eq:#1})}
\def\am{\left(\begin{array}{c}}
\def\amm{\left(\begin{array}{cc}}
\def\ammm{\left(\begin{array}{ccc}}
\def\a{\end{array}\right)}
\def\ds{\displaystyle}
\def\Re{{\rm Re}}
\def\Im{{\rm Im}}

\title{The influence of complex material coverings on the bandwidth of antennas}

\author{S.A. Tretyakov$^1$, 
S.I. Maslovski$^1$,\\ A.A. Sochava$^2$, C.R. Simovski$^{1,3}$}

\date{$^1$Radio Laboratory~/~SMARAD,
Helsinki University of Technology\\
P.O. Box 3000, FIN-02015 HUT, Finland\\[2mm]
$^2$St.~Petersburg State Technical University\\
St.\ Petersburg, Russia\\[2mm]
$^3$State Institute of Fine Mechanics and Optics\\
St.\ Petersburg, Russia\\[4mm]
\today}

\begin{document}

\maketitle

\bigskip
\begin{abstract}

The influence of material coverings on the  antenna bandwidth is
investigated for antennas formed by thin electric or magnetic line
sources. It is shown that uniform thin layers of arbitrary passive
materials (including Veselago,  left-handed, or
double-negative materials) cannot help to overcome the
bandwidth limitations imposed by the amount of energy stored in the
antenna reactive field.
Alternative possibilities offered by complex composite materials in the
antenna design are identified.

\end{abstract}

\newpage

\section{Introduction}

Properties of antennas depend on materials that may cover metal radiating
parts or fill the antenna volume.
For instance, it is known that the bandwidth of
microstrip antennas can be improved  by using magnetic
substrates \cite{Hansen,Olle,Ziol1}.
In the recent literature
there is renewed interest to the question of
improving antenna performance by loading the antenna with
complex materials.
In particular, it is of interest if new artificial materials with
negative material parameters \cite{negative} can be useful in the antenna design.
This problem has been analyzed in a recent conference presentation
\cite{Ziol2} for a point dipole antenna surrounded by a shell
of a material with negative parameters, and some very promising
results reported. However, the key issue of the reactive energy
storage and the antenna quality factor calls for a more careful
examination, because of inevitable frequency dispersion
of passive artificial materials.

In this paper we analyze in detail a very simple
radiating system: an infinite current line in a material cylindrical
shell. Assuming the diameter to be small compared with the
wavelength in the material, the field equations can
be solved analytically leading to simple and physically clear formulas.
The quality factor can be then defined and calculated for
arbitrary dispersive and also for specific lossy coverings.
The results of the analysis lead to some general conclusions
regarding the effect of thin material coverings on the antenna
performance.

\section{Fields of a thin radiating material-covered cylinder}

\begin{figure}[ht]
\centering
\epsfig{file=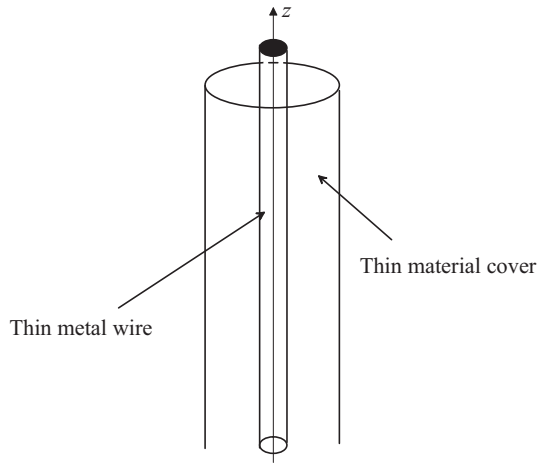, width=7cm}
\caption{Radiating current line in a thin material cylinder.}
\label{geom}
\end{figure}

We consider the radiation from a line electric or magnetic current
embedded into a material cylinder (Figure~\ref{geom}). The problem
originates from a simplified model of a thin-wire antenna
covered by a thin layer of some material. The antenna bandwidth is investigated by means
of the effective impedance per unit length of the antenna wire.
The real part of the introduced effective impedance corresponds to
the radiated power (radiation resistance), and the imaginary part
gives information about the energy stored in the antenna near
field. A thin material layer covering the radiating current
affects the stored energy and, if the material is lossy, the
radiated power. Our aim is to find how this changes the antenna
bandwidth.

Let us consider an infinitely thin line of time-harmonic electric
current $I$. We assume the current to be uniform in its phase and
its amplitude along the line. We model a thin radiating wire of
radius $r_0$ as such current line sitting at the wire axis. This
method is similar to taking into account only the local (singular)
part of the thin-wire kernel (e.g., \cite{Halen}). We denote  the
covering cylinder outer radius as $a$, the frequency of operation
as $\omega$, and the material parameters of the outer space and
the covering as $\E_0$, $\M_0$, and $\E$, $\M$, respectively.
Hence, for the wave number in the outer space we have $k_0 =
\omega\sqrt{\E_0\M_0}$, and for that in the material we have $k =
\omega\sqrt{\E\M}$. The branch of the square root in the last
relation is chosen so that $\Im\{k\} \le 0$ (the time dependence
is of form $e^{+j\omega t}$).

Expressing the line source fields inside and outside the cover and
using the assumed cylindrical symmetry we write for the electric
field: \e \cases {E_z= A H^{(2)}_0(kr)+B H^{(1)}_0(kr),\ \ \
\hfill r_0\le r \le a, \cr E_z=C H^{(2)}_0(k_0r),\hfill r\ge a,
\cr} \f and for the magnetic field: \e
H_{\varphi}=-{j\over{\omega\M}}{\partial E_z\over{\partial r}} =
\cases {\ds {jk\over{\omega\M}}\left(A H^{(2)}_1(kr)+B
H^{(1)}_1(kr)\right), \ \ \ \hfill r_0\le r \le a \cr \ds
{jk_0\over{\omega\M_0}}C H^{(2)}_1(k_0r),\hfill r\ge a \cr}. \f
Here the $z$-axis is directed along the wire and $r$ is the radial
distance counted from the axis of the wire [$(r,\varphi,z)$ form a
usual set of cylindrical coordinates].

On the surface $r=a$ we have the continuity boundary conditions
for $E_z$ and $H_{\varphi}$ field components: \e A H^{(2)}_0(ka)+B
H^{(1)}_0(ka)=C H^{(2)}_0(k_0a), \l{firstE} \f \e {\M_0 k\over{\M
k_0}}\left(A H^{(2)}_1(ka)+B H^{(1)}_1(ka)\right)=C
H^{(2)}_1(k_0a). \l{secondH} \f In addition, we use a relation
between the wire current $I$ and the magnetic field $H_{\varphi}$
at the wire surface $r=r_0$: \e 2\pi r_0 H_{\varphi} = I, \f
obtaining \e {2\pi j\, kr_0\over{\omega\M}}\left(A H^{(2)}_1(kr_0)
+ B H^{(1)}_1(kr_0)\right)=I. \l{thirdI} \f Eqs.~\r{firstE},
\r{secondH}, and \r{thirdI} form a complete system of equations.
Instead of solving this system directly in terms of Hankel
functions we make a simplification assuming that the following
thin-wire condition is satisfied: $kr_0 \ll 1$. Under this
assumption \e H^{(1,2)}_1(z)\approx \mp {2j\over \pi z}\ , \f and
Eq.~\r{thirdI} becomes \e B = A + {\omega \M \over 4}I. \l{B} \f
We substitute the last relation into Eqs.~\r{firstE} and
\r{secondH} to obtain \e 2J_0(ka)A - H_0^{(2)}(k_0a)C =
-{\omega\M\over 4}IH_0^{(1)}(ka), \f \e 2J_1(ka)A - {\M
k_0\over\M_0 k}H_1^{(2)}(k_0a)C = -{\omega\M\over
4}IH_1^{(1)}(ka). \f Finally, we get the solution in the following
form: \e A=-{\omega\M I \over{8}}{H^{(1)}_1(ka)H^{(2)}_0(k_0a)-\ds
{\M k_0\over\M_0 k}
H^{(1)}_0(ka)H^{(2)}_1(k_0a)\over{J_1(ka)H^{(2)}_0(k_0a) -\ds{\M
k_0\over\M_0 k} J_0(ka)H^{(2)}_1(k_0a)}}\ , \f \e C=-{j\omega\M
I\over{4}}{J_0(ka)Y_1(ka)-J_1(ka)Y_0(ka)\over{\ds
J_1(ka)H^{(2)}_0(k_0a) -\ds {\M k_0\over\M_0 k}
J_0(ka)H^{(2)}_1(k_0a)}}\ . \f The last unknown $B$ can be found
from Eq.~\r{B}.

The obtained solution is quite  difficult to analyze in the
general case. However, since we are interested in the material
effects in the case of thin coverings, when the antenna cross
section remains small in terms of the wavelength, we can assume
that $k_0a \ll 1$, $|k|a \ll 1$. Then, using the known Bessel
function asymptotics for small arguments \e J_0(z) \approx 1\ ,
\quad J_1(z) \approx {z\over 2}\ , \quad Y_0(z) \approx {2\over
\pi}\log{\gamma z\over 2}\ , \quad Y_1(z) \approx -{2\over \pi z}\
, \f (here $\gamma \approx 1.781$ is the Euler constant), the
relations for amplitudes $A$, $B$ and $C$ reduce to \e A =
-{\omega I\over 8}\left\{\M_0 + \M +
{2j\over\pi}\left[\M\log{\gamma k a\over 2} - \M_0\log{\gamma k_0
a\over 2}\right]\right\}, \l{approxA} \f \e B = -{\omega I\over
8}\left\{\M_0 - \M + {2j\over\pi}\left[\M\log{\gamma k a\over 2} -
\M_0\log{\gamma k_0 a\over 2}\right]\right\}, \l{approxB} \f \e C
= -{\omega\M_0 I\over 4}. \l{approxC} \f In the next section we
will use Eqs.~\r{approxA}--\r{approxC} to find the energy stored
in the source reactive field.

\section{Antenna quality factor}

The  antenna bandwidth is related with the antenna quality factor.
The well-know definition for the quality factor of a lossy (in the
antenna case losses are mostly due to radiation) resonator is \e Q
= {\omega W \over P}, \l{quality} \f where $W$ is the average
reactive energy stored in the resonator, and $P$ is the total
dissipated power (for an antenna this is a sum of the radiated
power and the thermal loss power). Although $Q$ defined by this
formula is mostly used for resonant systems, it can be also used
as a relative measure of ability of an aperiodic system to store
reactive energy.

Usually, the quality factor of antennas is evaluated integrating
the reactive field energy density over the whole space and
integrating the Poynting vector in the far zone. In our present
case, where  both the stored energy and the radiated energy are
due to the current in the antenna wire, we can use a simpler
method introducing the effective impedance per unit length of the
wire as follows: \e Z = -{E_z(r_0)\over I}, \l{impdef} \f where
$E_z(r_0)$ means the total longitudinal electric field component
created by the considered line current at the surface $r=r_0$. The
minus sign is because we suppose $I$ to be given (external)
current. This approach is sometimes called the induced
electromotive force method. The field component of interest is
given by \e E_z(r_0)= A H^{(2)}_0(kr_0) + B H^{(1)}_0(kr_0). \f
Using the thin-wire approximation we express this as \e E_z(r_0) =
A\left(1 - {2j\over \pi}\log{\gamma k r_0\over 2}\right) +
B\left(1 + {2j\over \pi}\log{\gamma k r_0\over 2}\right). \f
Substituting $A$ and $B$ from Eqs.~\r{approxA}, \r{approxB}, and
using Eq.~\r{impdef} we get \e Z = {\omega\M_0\over 4} +
j{\omega\over 2\pi}\left[\M\log{a\over r_0} + \M_0\log{4\over
\gamma k_0 a}\right]={\omega\M_0\over 4} + j{\omega\over
2\pi}\left[\M_0\log{4\over \gamma k_0r_0} + (\mu-\M_0)\log{a\over
r_0}\right]. \l{Z_antenna} \f This formula has a clear physical
meaning. ${\omega \mu_0 \over 4}$ is the radiation resistance per
unit length. The first member in the brackets is the reactance per
unit length in the absence of covering. The last member,
proportional to $\mu-\mu_0$, measures the influence of the
covering on the reactive input impedance. Note that under our
assumptions the radiation resistance does not depend on the
parameters of the thin covering cylinder.

The obtained impedance is a complex number: $Z(\omega) = R(\omega)
+ jX(\omega)$, where \e R(\omega) = {\omega\M_0\over 4} +
{\omega\M''(\omega)\over 2\pi}\log{a\over r_0}\ , \l{ReZ} \f \e
X(\omega) = {\omega\over 2\pi}\left[ \M_0\log{4\over \gamma k_0
r_0}+ (\M'(\omega)-\mu_0)\log{a\over r_0} \right]. \l{ImZ} \f Here
we have assumed loss only in the covering material: $\M = \M' -
j\M''$.

\subsection{Negligible losses in the covering material}
\label{lossless}

Let us first assume that the losses in the material cover can be
neglected. In this case we can find the quality factor in terms of
the derivative of $X(\omega)$ with respect to the frequency. Here,
we follow the approach used in \cite{Vainshtein}. Instead of
considering steady-state harmonic oscillations, we study a
transient regime characterized by an exponentially growing (or
decaying) amplitude of harmonic oscillations. For this regime we
introduce a complex frequency: \e \Omega = \omega - j\alpha. \f
Positive values of $\alpha$ result in growing oscillations,
negative values correspond to decaying ones. We assume that
$|\alpha| \ll \omega$. This implies a very slow rate of the
amplitude change, hence, for the reactive energy $W$ (averaged
over a period of oscillations) stored in the circuit  we can write
\e {\d W\over \d t} \approx 2\alpha W. \l{energy} \f The
coefficient 2 is there because the energy is proportional to the
square of the oscillation amplitude. The circuit impedance as a
function of the introduced complex frequency becomes \e Z(\Omega)
= R(\Omega) + j X(\Omega) \approx R(\omega) + \alpha{\d
X(\omega)\over \d \omega} + j\left[X(\omega) - \alpha {\d
R(\omega)\over \d \omega}\right]. \f We see that exponential
growth of oscillation amplitude results in an additional
resistance as well as an additional reactance.

In the case of negligible dissipation loss which we consider at this stage
it is possible to split the total active power
(proportional to the real part of the effective impedance) into
parts corresponding to radiation and increase of the
stored reactive energy. Indeed, the radiated power can be
calculated knowing the far-zone fields. If there is no dissipation loss
in the antenna, then the way how to find the energy storage power is
straightforward. A complexity comes when the loss and storage
processes are combined (see the next section).

Comparing Eqs.~\r{approxC} and \r{ReZ} we see that for our case
the radiation resistance (per unit length) is \e R_{\rm rad} =
{\omega\M_0\over 4}. \f Then the energy storage power is given by
(we use the effective values of the fields and the current) \e {\d
W\over \d t} = \left(\Re\{Z(\Omega)\} - R_{\rm rad}\right)|I|^2 =
\alpha{\d X(\omega)\over \d \omega}|I|^2. \f Comparing this with
Eq.~\r{energy} we find the averaged stored reactive energy (per
unit length of the wire) \e W={1\over 2}{\d X(\omega)\over \d
\omega} |I|^2 \ , \f and obtain for the quality factor: \e Q =
{\omega W \over P} = {\omega\over 2 R_{\rm rad}}{\d X(\omega)\over
\d\omega} \ , \l{QX} \f because the total loss power is the
radiated power $P = R_{\rm rad} |I|^2$. Differentiating
$X(\omega)$ given by \r{ImZ}, we get \e {\d X(\omega)\over
\d\omega} = {1\over 2\pi}\left\{{\d(\omega \M)\over
\d\omega}\log{a\over r_0} + \M_0\left[\log{4\over\gamma k_0 a } -
1\right] \right\}. \f We substitute the last result into
Eq.~\r{QX} and do simple algebra to get finally \e Q = {1\over
\pi}\left\{{\d(\omega \M_r)\over \d\omega}\log{a\over r_0} +
\log{4\over\gamma k_0 a} - 1 \right\}, \l{Qfinal} \f where we have
introduced $\M_r = \M/\M_0$ as the relative permeability of the
covering material.

\subsection{Lossy and dispersive covering cylinder}

If the covering material is dispersive and the dissipation losses
cannot be neglected, the quality factor cannot be determined
from the analysis of the antenna impedance only. The reason is that
the active part of the impedance determines
the {\em total} active power delivered to the antenna, but we need
to know what part of this power is used to increase the stored energy and
what part is dissipated into heat.
The quality factor can be, however, found, if we actually know  the
internal structure of the analyzed system. In our particular case,
the result will depend on  the structure and design of the loading material.

\begin{figure}[ht]
\centering
\epsfig{file=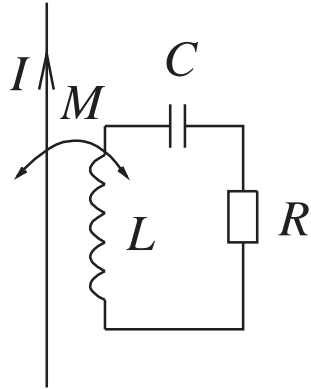, width=4cm}
\caption{Equivalent circuit for a unit-length section of a
wire antenna covered by a lossy and dispersive
magnetic material
cylinder.}
\label{circuit}
\end{figure}

Let us assume that the magnetic material that fills the covering
cylinder can be modeled by a Lorentzian permeability: \e
\mu=\mu_0\left(1+{A\omega^2\over{\omega_0^2-\omega^2+j\omega
\Gamma}}\right), \l{Lorentz}\f where the magnitude factor $A$ and
the loss factor $\Gamma$ do not depend on the frequency. This
material can be realized, for example, as a dense array of small
split-ring resonators made of metal wires or strips. The total
impedance of the covered antenna \r{Z_antenna} we write as a sum
of the antenna impedance without material cover and an additional
impedance due to the material cylinder: \e Z=Z_{\rm wire}+Z_{\rm
medium} \f with \e Z_{\rm wire}={\omega\M_0\over 4} +
j{\omega\mu_0\over 2\pi}\log{4\over \gamma k_0r_0},\qquad
 Z_{\rm medium}= j{\omega\over
2\pi}(\mu-\M_0)\log{a\over r_0}.
\l{impedances}
 \f
If the permeability obeys \r{Lorentz}, to model $Z_{\rm medium}$
we can introduce a magnetic-coupled circuit shown in
Figure~\ref{circuit}. Indeed, in terms of the introduced effective
parameters, the additional impedance associated with the material
can be written as \e Z_{\rm material} = { j\omega^3 (M^2/L) \over
\omega_0^2 - \omega^2 +j\omega (R/L)} \f (here $\omega_0 =
1/\sqrt{LC}$) which has the same form as the impedance following
from \r{impedances} and \r{Lorentz}. Given the total current
through the antenna wire $I$, we can now find the currents and
voltages on the reactive elements of the equivalent circuit and
calculate the averaged stored energy as \e W_{\rm medium}
=L{|I_L|^2\over 2} + C{|U_C|^2\over 2}= {1\over 2}
\left(L+{1\over{\omega^2 C}}\right)|I_L|^2 .\f The result is \e
W_{\rm medium}={\omega^2M^2C(1+\omega^2 LC) \over(1-\omega^2LC)^2
+ \omega^2R^2C^2}{|I|^2\over 2}. \f At the resonant frequency the
last expression becomes \e W_{\rm medium}(\omega_0) =
{M^2|I|^2\over R^2C}. \f We observe that if there is no coupling,
the additional energy stored in the medium is zero, as it should
be. The dissipated power is \e P_{\rm loss} = R |I_L|^2
={\omega^4M^2C^2R|I|^2\over (1-\omega^2LC)^2 + \omega^2R^2C^2}\f
At the resonant frequency the dissipation is given by \e P_{\rm
loss}(\omega_0) = {\omega_0^2M^2|I|^2\over R}. \f The radiation
efficiency of the considered system at the resonant frequency can
be found as \e \eta = {P_{\rm rad}\over P_{\rm rad} + P_{\rm
loss}}  = {1\over \ds 1 + {4\omega_0 M^2\over \M_0 R}}, \f and the
total antenna quality factor is \e Q_{\rm total} = {\omega (W_{\rm
wire} + W_{\rm medium})\over P_{\rm rad} + P_{\rm loss}} =
\eta{\omega (W_{\rm wire} + W_{\rm medium})\over P_{\rm rad}}. \f
Finally, \e Q_{\rm total} = {1\over \ds 1 + {4\omega_0 M^2\over
\M_0 R}}\left[{1\over \pi} \left(\log{4\over\gamma k_0
r_0}-1\right) +{4M^2\over\M_0 R^2C}\right]. \l{totalQ} \f

Naturally, one wants to minimize dissipation in the antenna in
order to increase the antenna efficiency. However, from \r{totalQ}
it is seen that if the losses in the material become very small
($R \rightarrow 0$), the total quality factor reduces to the
well-known relation for an oscillatory circuit: \e Q_{\rm total} =
{1\over \omega_0 RC}, \f which means that most of the energy is
stored in the medium layer and almost all the source power goes
into heat (the antenna efficiency tends to zero).

Next, let us fix a given efficiency value fixing the resistance,
the resonant frequency and the coupling. To decrease \r{totalQ}
one should increase the capacitance. In the limit $C \rightarrow
\infty$. For a given fixed resonant frequency that means $L
\rightarrow 0$. One can see that the total impedance of a medium
particle and the medium effective impedance become simply
resistive in the limiting case: \e Z_{\rm medium} =
{\omega_0^2M^2\over R}. \f Hence, the deepest drop in the total
quality factor can be achieved if the medium loading is equivalent
to a simple resistor. The limiting value for the quality factor is
for this case \e Q_{\rm total} = \eta Q_{\rm wire}, \f as
expected. Note also that when the inductance $L$ tends to zero, a
better model should assume that also the mutual inductance $M$
tends to zero. In that case the effect due to the material cover
simply disappears in the limit, as the particles forming the cover
material are not excited.

Because the radiation resistance of the considered
system does not depend on the layer parameters and, moreover, the
covering does not affect the part of energy stored {\it outside}
the cylinder, the antenna quality factor and the bandwidth {\it
associated with radiation} does not change in the presence of
covering. However, the total quality factor associated with the
input impedance of the antenna as a circuit element is very much
affected.

\section{Discussion and conclusions}

First of all, we see from Eq.~\r{Qfinal} that the electric
properties of the covering material have disappeared from the
antenna quality factor expression. Such situation is usual when
one considers thin coverings placed on top of electric conductors.
But a more important conclusion following from Eq.~\r{Qfinal} is
that we cannot achieve any improvement in the antenna bandwidth
even covering it with a (passive) magnetic material having a
negative permeability. This follows from the causality limitations
on the material parameters of any  passive media with negligible
losses \cite{Landau}: \e {\d(\omega\E_r)\over \d\omega} \ge 1,
\quad {\d(\omega\M_r)\over \d\omega} \ge 1. \f Indeed, because of
these limitations, the value of the first term in braces in
Eq.~\r{Qfinal} can only be increased compared to the case of empty
filling (i.e. when $\M = \M_0$), in turn increasing $Q$ and
narrowing the bandwidth.

Considering a dual problem involving a magnetic current line
as the radiating source covered with the same material, one will
get the same expression \r{Qfinal}, only the magnetic permeability will
be replaced by the electric permittivity. There will be no bandwidth improvement also.

In case of our example of an antenna with a dispersive and lossy covering,
the additional stored reactive energy is of course also always positive,
although the reactive part of the input impedance can cross zero and the
derivative of $X(\omega)$ can change sign. The bandwidth can increase
due to additional loss at a cost of reduced antenna efficiency.
The formulas for the stored energy in  this case are not universal and
depend on the material structure, but the conclusion that the stored
energy is positive in any passive system is universal.

To summarize, homogeneous coverings of radiating metal wires\footnote{For finite-length
wire antennas, the present theory is applicable to covers that do not
substantially change the effective antenna height.} by
{\em electrically thin}
passive material layers lead to reduced bandwidth or lower efficiency,
whatever exotic properties these covering materials might have.
The following alternative possibilities to improve antenna performance using
complex materials can be identified:

\begin{itemize}

\item The use of {\em radiating} inclusions, and
not really {\em material} coverings.

\item Shells of resonant dimensions. New materials can offer more
possibilities in optimizing resonant antennas.
Apparently, this is the case considered in \cite{Ziol2}.

\item Nonuniform coverings or material inclusions. This can
modify the current distribution, possibly leading to
increased bandwidth.

\item Active materials. If the passivity requirement is dropped,
it is in principle possible, for example, to realize wide-band weakly dispersive
negative material parameters \cite{negative_wide}.
The stored reactive energy can be negative
(and bandwidth very large),
meaning, effectively, that the whole volume filled by this material
is the source of power.

\end{itemize}


\begin{thebibliography}{99}


\bibitem{Hansen}
R.C. Hansen, M. Burke, Antennas with magneto-dielectrics,
{\it Microwave and Optical Technology Letters,} vol. 26, no. 2, pp. 75-78, 2000.


\bibitem{Olle}
O. Edvardsson, On the influence of capacitive and inductive loading on different types of small
patch/PIFA structures for use on mobile phones,
{\it 11th International Conference on Antennas and Propagation,}
vol. 2, pp.\ 762-765, April 2001.


\bibitem{Ziol1}
S. Yoon and R.W. Ziolkowski,
Bandwidth of a microstrip antenna on a magneto-dielectric substrate,
{\it IEEE Antennas and Propagation Symposium,} Columbus, Ohio, June 22-27, 2003.


\bibitem{negative}
D.R. Smith, W.J. Padilla,
D.C. Vier, S.C. Nemat-Nasser, and S. Schultz, Composite media with simultaneously
negative permeability and permittivity,
{\it Physical Review Lett.,} vol.\  84,
pp.\ 4184-4187, 2000.





\bibitem{Ziol2}
R.W. Ziolkowski and A.D. Kipple,
Application of double negative materials
to modify the performance of electrically small antennas,
{\it IEEE Antennas and Propagation Symposium,} Columbus, Ohio, June 22-27, 2003.



\bibitem{Halen}
C.A. Balanis, {\em Antenna Theory: Analysis and Design,} 2nd ed.,
Chapter 8,
N.Y.: John Wiley \& Sons, 1997.




\bibitem{Vainshtein}
L.A. Vainshtein, {\em Electromagnetic Waves,} Moscow: Radio i Sviaz, 1988 (in Russian).

\bibitem{Landau}
L.D. Landau and E.M.   Lifshits,  {\it Electrodynamics of Continuous Media,}
2nd ed.,  Oxford, England: Pergamon Press, 1984.


\bibitem{negative_wide}
S.A. Tretyakov, Meta-materials with wideband negative
permittivity and permeability, {\it Microwave and Optical Technology Letters,}
vol. 31, no. 3, pp. 163-165, 2001.


\end{thebibliography}
\end{document}